\documentclass[prl,aps,twocolumn,groupedaddress,floats,showpacs,final,superscriptaddress]{revtex4}
\usepackage{graphicx}
\usepackage{bm}
\usepackage[pdftex]{hyperref} 

\begin{document}

\author{Andrey S. Mishchenko}
\affiliation{RIKEN Center for Emergent Matter Science (CEMS),
2-1 Hirosawa, Wako, Saitama, 351-0198, Japan}
\affiliation{Russian Research Center ``Kurchatov Institute'', 123182 Moscow, Russia}
\author{Naoto Nagaosa}
\affiliation{Center for Emergent Matter Science, RIKEN (CEMS),
2-12-1 Hirosawa, Wako, Saitama, 351-0198, Japan}
\affiliation{Department of Applied Physics, The University of Tokyo,
7-3-1 Hongo, Bunkyo-ku, Tokyo 113, Japan}
\author{Nikolay Prokof'ev}
\affiliation{Department of Physics, University of Massachusetts, Amherst, MA 01003, USA}
\affiliation{Russian Research Center ``Kurchatov Institute'', 123182 Moscow, Russia}


\title{ Diagrammatic Monte Carlo method for many-polaron problems }
\date{\today}


\begin{abstract}
We introduce the first bold diagrammatic Monte Carlo approach to deal with
polaron problems at finite electron density non-perturbatively, i.e.,
by including vertex corrections to high orders.
Using Holstein model on a square lattice as a prototypical example, we demonstrate that our
method is capable of providing accurate results in the thermodynamic limit in all regimes
from renormalized Fermi-liquid to a single polaron, across the non-adiabatic region where
Fermi and Debye energies are of the same order of magnitude. By accounting for vertex corrections
the accuracy of theoretical description is increased by orders of magnitude relative to the
lowest-order self-consistent Born approximation employed in most studies.
We also find that for the electron-phonon coupling typical for real materials,
the quasiparticle effective mass increases and the quasiparticle residue decreases
with increasing the electron density at constant electron-phonon coupling strength.
\end{abstract}


\maketitle


The nature of a solid state material implies existence of the electron-phonon
interaction (EPI) originating from Coulomb interaction between electrons
and nuclei. Its treatment is a notoriously difficult
non-perturbative task. In the limit of vanishing electron density we have to deal with
polarons defined as quasiparticle states with properties significantly
(sometimes radically) modified relative to the ``bare" particle states by their
interaction with the environment. Polarons are found across all fields in physics with
the same questions about their dispersion relation, effective mass, quasi-particle residue,
etc. being asked for different types of particles, environments and coupling between
them \cite{Landau,Appel}.

At large electron density, we face the many-polaron problem, or generic
interacting system of electrons and phonons. The standard simplified procedure to deal with
EPI in Fermi liquids (FL) is based on the adiabatic approximation taking advantage of the small ratio
$\gamma=\omega_D/\epsilon_F \sim \sqrt{m_e/m_i} \ll 1$ between the Debye, $\omega_D$, and
Fermi, $\epsilon_F$, energies (here $m_{e,i}$ are the electronic and ionic masses, respectively).
It is assumed that all interactions between (and with) the heavy ions are instantaneously
screened by the static dielectric function of a metal,
and the phonon spectrum is determined from the corresponding dynamic matrix of a solid.
Thus transformed EPI is no-longer singular at small momenta but, nevertheless, remains strong
and does not involve natural small parameters in realistic materials;
the dimensionless coupling constant $\lambda$ (to be defined below) is of the order of unity.
[Needless to say that EPI has to be strong enough to mediate $s$-wave superconductivity
in a system with Coulomb repulsion.] Such FL parameters as $Z$-factor and effective mass $m^*$
are still controlled by $\lambda \sim 1$. To leading order, this physics is adequately captured
by the self-consistent Born approximation (non-crossing self-energy diagrams) because according
to Migdal's theorem \cite{Migdal} vertex corrections are small in the adiabatic parameter $\gamma$.
Calculating vertex corrections precisely remains a daunting task. Still, it has to be completed in order to
(i) establish accuracy limits of the leading approximation, (ii) improve precision of
the theoretical description, and (iii) describe cases with $\gamma \sim 1$, intermediate between
the FL and single-polaron regimes.

In this work, we address the many-polaron problem within the
Holstein model \cite{Holstein} on a square lattice
\begin{equation}
H=-t \sum_{<i,j>} c^{\dagger}_{i}c_{j}^{\,} +\omega_0
\sum_{i}b_{i}^{\dagger} b_{i}^{\,} + g \sum_{i} c_{i}^{\dagger
}c_{i}^{\,} \left( b_{i}^{\dagger} +b_{i}^{\,} \right) \;.
\label{h0}
\end{equation}
We employ standard notations for electron/phonon creation (and annihilation) operators
$c_{i}^{\dagger}$/$b_{i}^{\dagger}$. Here $t$ is the nearest neighbor hopping amplitude,
$\omega_0$ is the energy of the local optical mode, and $g$ is the strength of EPI 
[a convenient dimensionless parameter is $\lambda = g^2 / (4 \omega_0 t)$].
In what follows the lattice constant $a$, amplitude $t$, and
Planck's constant $\hbar$, are used as units of length, energy, and time, respectively.
In the single polaron limit the crossover from weak- to strong-coupling regimes occurs
at $\lambda_c \approx 1$. The formation of the bipolaron bound state
is predicted to occur at $\lambda \approx 0.5$ \cite{Macridin}.

A consistent theory of EPI cannot be separated from Coulomb forces
between the electrons. Indeed, acoustic phonons in metals
do not even exist in the absence of EPI since their energies are shifted all the way up to
the ionic plasma frequency. Once electron-phonon and Coulomb interactions
between the electrons are accounted for, the acoustic spectrum is recovered back due to
screening of the long-range forces \cite{Brovman}. When further progress is made by separating
effects of EFI and electron-electron interactions \cite{Migdal,Eliashberg},
double-counting is dealt with approximately by excluding static electronic polarization
from the renormalization of phonon propagators. As a result, the effects of EPI on
crystal vibrations turn out to be small in the adiabatic parameter $\gamma$, but
accounting for the remaining terms in the phonon self-energy after that is,
strictly speaking, an ill-defined procedure in the absence of electron-electron interactions.
Given this subtlety, numerous work simply neglects all effects of
EPI on the phonon subsystem, and here we follow the same route (for the most part).


Last two decades have seen a remarkable progress in developing
unbiased numerical methods for a single polaron \cite{PS98,MPS2000,Kornilovitch,trugman}
(see also Ref.~\cite{Pola1,Pola2} for recent reviews).
While these methods provide extremely accurate description of nearly all aspects of
the polaron physics, in their present form none is suitable for performing precise
calculations for finite-density systems all the way to FL with $\gamma \ll 1$.  
There also exist numerous studies of dense polaron systems in a rather special 
one dimensional (1D) case.\cite{1d1,1d2,1d3,1d4,1d5}

Studies of 2D Holstein \cite{DMFT_h1,DMFT_h2} and Holstein-Hubbard models
\cite{DMFT_hh1,DMFT_hh2,DMFT_hh3} are limited to dynamical mean-field theory (DMFT) and
Determinant Monte Carlo (DMC) \cite{Scal_01,Scal_02,Scal_03,Scal_04,Scal_05} approaches.
The latter method was the first successful attempt to address EPI problems
in the many-body set-up systematically.
Being free from systematic errors, it is, however, not completely generic because it
faces the sign-problem for spin-imbalanced systems and
non-local interactions, and is computationally expensive for large system sizes. These problems
are absent in DMFT at the expense of unknown systematic error coming from the assumption that
the electron self-energy is purely local.
In comparison, our method is generic: it can treat systems in the thermodynamic limit with non-local
interactions and dispersive phonons (e.g. acoustic modes)
at arbitrary chemical potentials for both spin components and, importantly, provides
estimates for systematic error bars on final answers. The last feature is a crucial step
towards controllably accurate description of EPI required for material science.

The solution described in this work is based on the Bold Diagrammatic Monte Carlo (BDMC) technique
\cite{bold1,NaturePhys,spins} that takes advantage of field theoretical methods
to compute skeleton (irreducible and fully renormalized) free-energy diagrams
to high orders using stochastic sampling protocols. By applying BDMC to solve
Eq.~(\ref{h0}) at finite chemical potential and temperature on a square lattice,
we observe that the skeleton series converge not only in
the single-polaron and FL limits (as expected) but also in the non-adiabatic
parameter regime when $\gamma \sim 1$.
By accounting for vertex corrections, the accuracy of theoretical description
is radically improved from about 5\% (for the lowest-order treatment)
down to 0.2\%. Contrary to expectations that quasiparticle properties
are most strongly renormalized in the single polaron limit, we find that
the effective mass increases and the quasiparticle residue decreases with
increasing the electron density at constant EPI.
 
Our implementation of BDMC is based on irreducible free-energy diagrams
in terms of exact propagators $G$ and $D$ for the electron and
phonon degrees of freedom, respectively, the so-called $G^2W$ expansion \cite{molinari}.
In close similarity with the BDMC formulation used for quantum spin models \cite{spins}
(the same updating scheme can used)
we expand the electronic self-energy $\Sigma^{(N)}$ and the polarization operator
$\Pi^{(N)}$ into series of irreducible skeleton graphs, up to order $N$ in the number of
$D$ propagators. Self-consistency is implemented by feedback loops when
$G$ and $D$ are obtained from the free propagators $G^{(0)}$ and $D^{(0)}$
by solving algebraic Dyson equations
$[G({\bf k},\omega_m)]^{-1}=[G^{(0)}({\bf k},\omega_m)]^{-1}-\Sigma^{(N)}({\bf k},\omega_m)$
and
$[D({\bf k},\omega_n)]^{-1}=[D^{(0)}({\bf k},\omega_n)]^{-1}-\Pi^{(N)}({\bf k},\omega_n)$
in momentum ${\bf k}$ and Matsubara frequency $\omega_m=2\pi T (m+1/2)$, $\omega_n=2\pi T n$
representation (here $m$ and $n$ are integer). The notorious sign-problem as we know it
(exponential increase of computational complexity with the system size) does not exist
in the space of connected Feynman diagrams that are formulated directly in the thermodynamic
(i.e. infinite system size) limit. Instead, sign-alternation of diagrams is a necessary condition(!)
for series convergence: with the number of diagrams of order $N$ increasing factorially,
cancelation of the same-order diagrams ensures that the BDMC technique produces converged
(or subject to re-summation methods) results. Establishing convergence properties of
the skeleton expansion for the EPI system is the most important methodological result of this work.
We refer studies of the superfluid, bipolaronic, etc. instabilities to future work and
thus limit ourselves here to the coupling constant $\lambda=0.45$ (just below the threshold for the bipolaron formation) and away from the nesting conditions at half-filling. Given that the bandwidth of the
tight-binding model is $W=8t$, we fix $\omega_0=0.5t$, low enough to guarantee that
we can reproduce the FL regime with $\gamma \ll 1$.
\begin{figure}[t]
\includegraphics[scale=0.4,width=0.9\columnwidth]{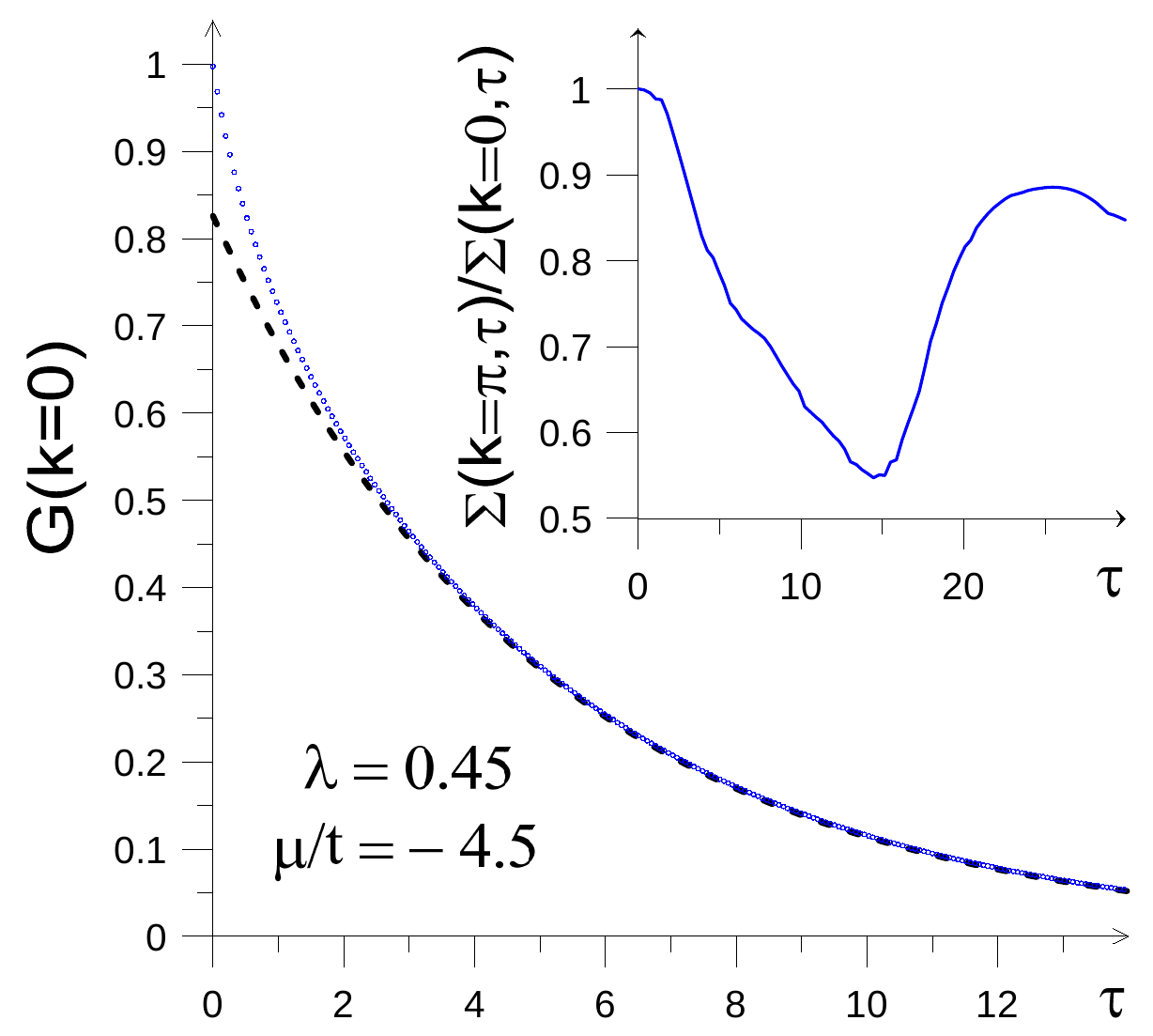}
\caption{\label{fig:1} (color online) Green's function of a single polaron
with zero momentum at $\lambda=0.45$, $\mu/t=-4.5$ and zero temperature
obtained by the BDMC technique (blue dots). It is compared to the
asymptotic behavior $G_{\tau \to \infty}(k=0)=Z_s \exp[-E_s \tau]$ (dashed line)
with $Z_s=0.826$ and $E_s=-0.302$ calculated by the conventional single-polaron
diagrammatic MC approach \cite{MPS2000}. All error bars are smaller than symbol sizes.
Inset: Ratio of the polaron self-energies $\Sigma({\bf k},\tau)$
at ${\bf k}=(\pi ,0)$ and ${\bf k}=0$ as a function of imaginary time $\tau$.}
\end{figure}

Feynman diagrams are typically formulated in the thermodynamic limit.
In practice, we consider a mesh of
$L^2=128^2$ momentum points with periodic boundary conditions, large enough
to ensure that results do not depend on $L$ within error bars.
\begin{figure}[t]
\includegraphics[scale=0.38,width=0.86\columnwidth]{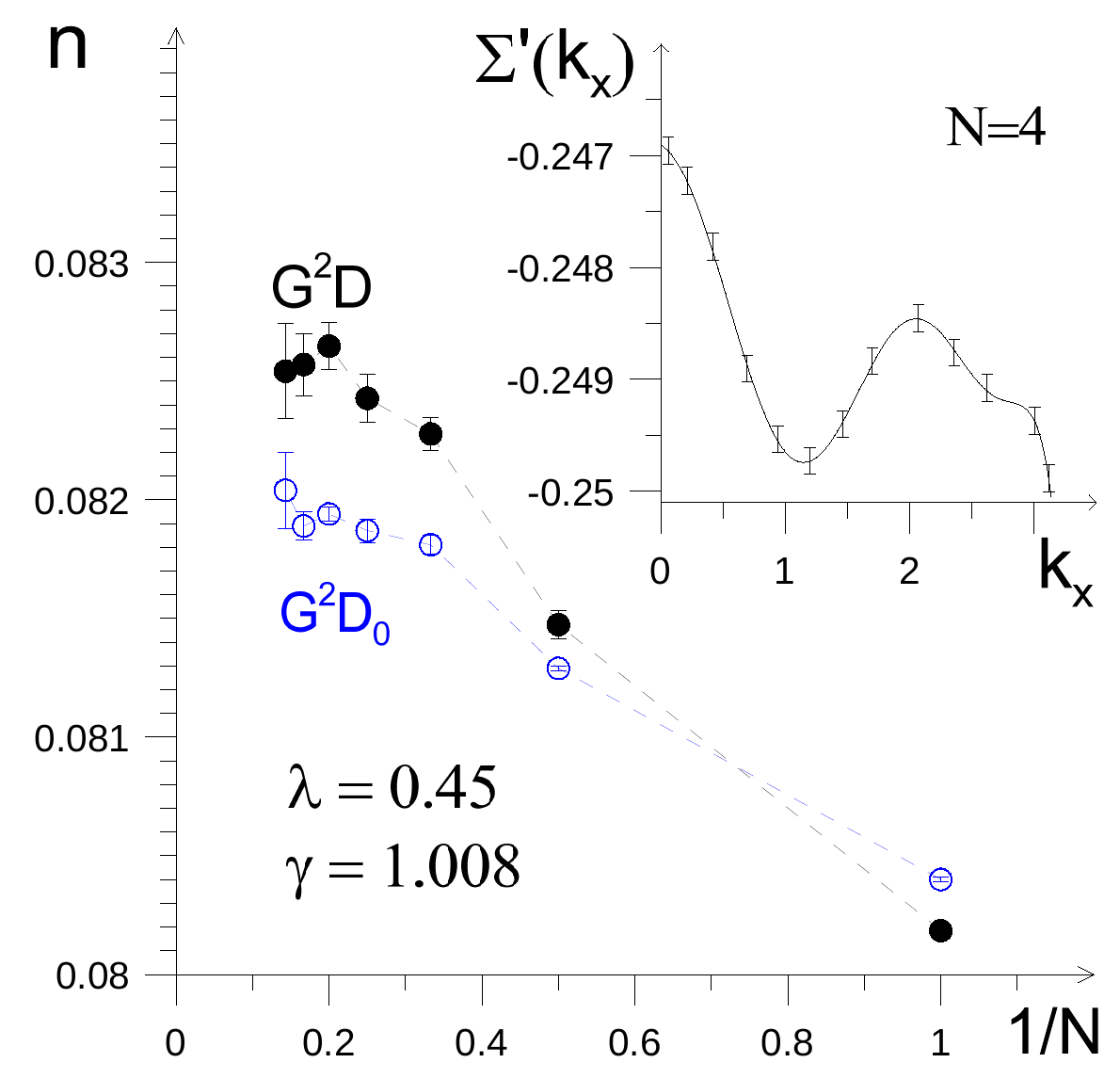}
\caption{\label{fig:2} (color online) Electron density computed from diagrams up to order $N$
without (open blue squares) and with (filled black circles) the
self-consistent renormalization of phonons at $\lambda=0.45$, $\gamma= 1.008$, and $T=0.01$.
We subtracted the static local polarization of the ideal electron gas
at the same density from the phonon self-energy to mimic the standard material science
protocol \cite{Brovman} where this contribution is already accounted for in the
value of $\omega_0$.
Inset: Momentum dependence of the static self-energy $\Sigma ({\bf k},0)$ along the $\hat{k}_x$ axis.}
\end{figure}

We start with establishing the convergence properties. In the single-polaron limit
the diagrammatic expansion is sign-positive and its convergence
is guaranteed. The difference between the more conventional diagrammatic approach
\cite{PS98,MPS2000} and BDMC is that all momenta are simulated in a single run and the
number of diagrams at any given order is reduced in the latter.
In this regime, one is not limited
by the maximum diagram order and all error bars are statistical in nature.
In Fig.~\ref{fig:1} we show that the BDMC approach perfectly reproduces known results
for a polaron. From the ratio of electron self-energies at momenta ${\bf k}=(\pi ,0)$ and ${\bf k}=0$,
see inset in Fig.~\ref{fig:1}, we conclude that $\Sigma ({\bf k},\omega_n)$ at finite frequencies
has appreciable momentum dependence while the static one is approximately ${\bf k}$-independent
[to reproduce known physics of end-points for a single polaron~\cite{Larsen},
$\Sigma$ has to have strong momentum dependence for excited states].

According to Migdal's theorem, we also expect convergence in the FL
regime (at least up to diagram orders comparable to $1/\gamma \gg 1$). It is thus crucial to study
the non-adiabatic regime $\gamma \sim 1$. In Fig.~\ref{fig:2} we present our data
for electron density at fixed chemical potential $\mu/t=-3.75$
(it corresponds to $\gamma \approx 1.008$)
as a function of the maximum diagram order accounted for in the simulation.
We observe converging behavior with most changes being exhausted by going from first- to
fourth-order diagrams;  all $2,017,881$ eighth-order diagrams  \cite{molinari} cancel each
other within the error bars. It is worth emphasizing here, that the lowest-order result
all by itself is meaningless despite the fact that it is capturing most of the answer because
its limits of accuracy can be established only through higher-order calculations.
In addition, Fig.~\ref{fig:2} makes it clear that the accuracy of the theoretical description
is improved at least by an order of magnitude (down to a fraction of a percent) if vertex corrections
up to forth-order are accounted for. Further improvements can be achieved only at the expense
of increased simulation time due to factorial growth of computational complexity with the diagram
order. The rest of the data presented in this work were obtained by performing simulations
up to fourth-order (with additional checks at selected points that six-order results remain
the same within the error bars). The inset in Fig.~\ref{fig:2} shows the $k_x$-dependence of the
static electron self-energy, where the momentum independence remains at the 2\% level,
similarly to that in Fig.~\ref{fig:1}. This outcome not only quantifies/validates the 
local DMFT approximation \cite{DMFT_h1,DMFT_h2,DMFT_hh1,DMFT_hh2,DMFT_hh3} 
for low-temperature thermodynamic properties but also the momentum-averaged approach used 
in Refs~\cite{Mona1,Mona2,Mona3,Mona4,Mona5}.

\begin{figure}[htbp]
\includegraphics[scale=0.4,width=0.88\columnwidth]{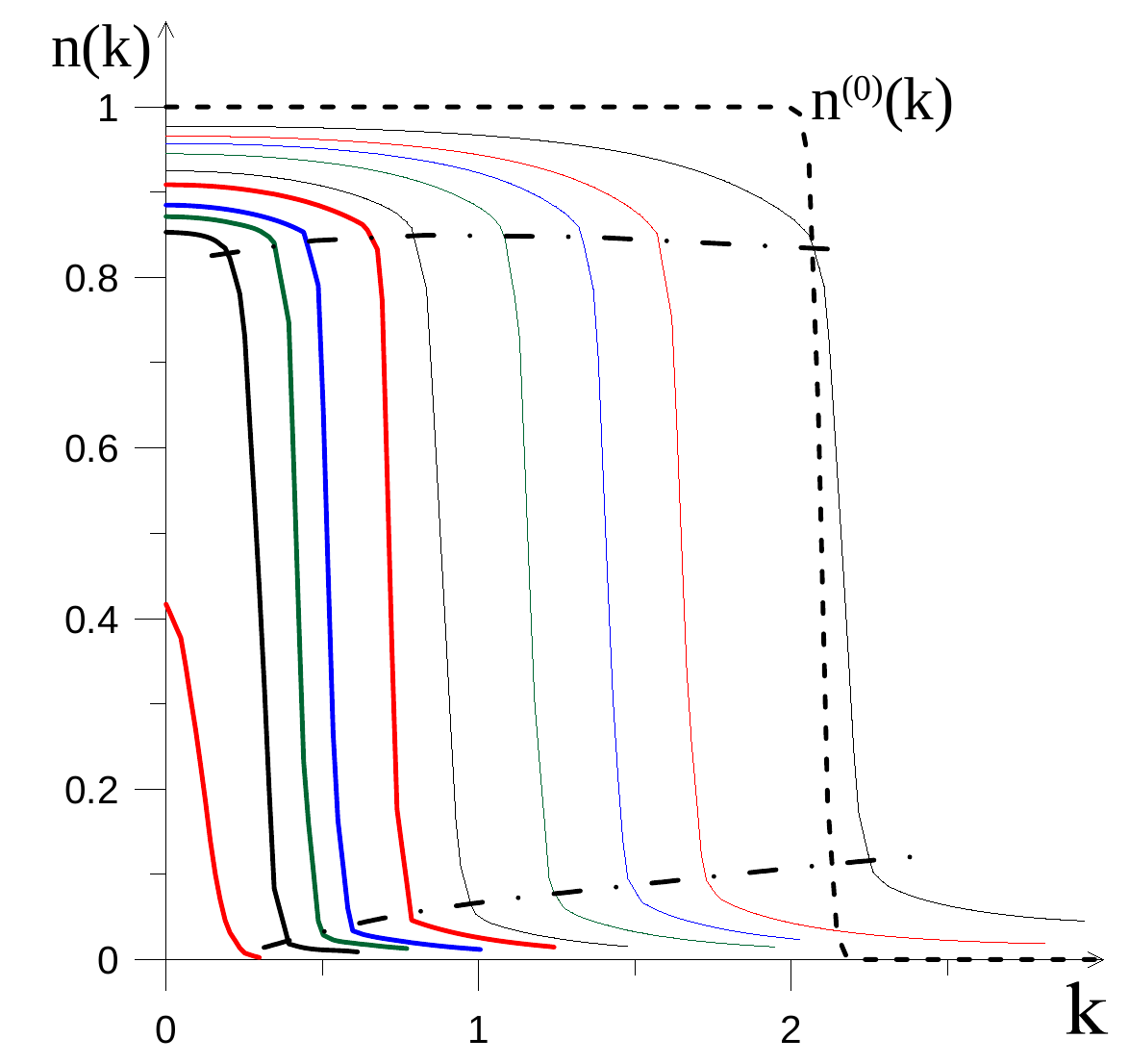}
\caption{\label{fig:3} (color online) Momentum distributions
for various electron densities per site (from right to left:
n=0.6498, \, 0.4055, \, 0.3026, \, 0.2086, \, 0.1222, \, 0.08169, \, 0.04323, \, 0.02846,
\, 0.01413, \, 0.001416 ) at $\lambda=0.45$. For $n>0.1$ we consider $T/t=0.025$ (thin lines); for
$n<0.1$ the temperature is reduced down to $T/t=0.01$ (thick lines).
The dashed line corresponds to the ideal gas case at $T/t=0.025$ and $\mu/t=-1$.
With dash-dotted lines we show (approximately) the locations of the
$T=0$ jumps in the distribution function.
}
\end{figure}

In Fig.~\ref{fig:3} we present the entire evolution of the electron momentum
distribution $n(k)$ from high ($ \gamma < 1/6$) to low ($\gamma \approx 6$) densities
with the characteristic jump at the Fermi momentum (smeared by finite-temperature effects).
At the lowest density the conventional Fermi-distribution transforms into the Gaussian
distribution characteristic of the dilute polaron gas at finite temperature $T=0.01t>\epsilon_F$.
By looking at Fig.~\ref{fig:3}, one might think that the quasiparticle residue $Z$
decreases from high to low densities, judging by the value of $1-n(k=0)$ and by invoking
an argument that Pauli-principle restrictions at the Fermi surface reduce the amount of
spectral weight transfer to incoherent continuum and quaisparticle ``dressing". This intuition
turns out to be completely wrong because one has to look at the discontinuity of the
distribution function at the Fermi surface in the limit of $T\to 0$ (see dashed-dotted lines
in Fig.~\ref{fig:3}).

To deduce the quasiparticle residue and effective mass at the Fermi surface we perform
standard data processing for the proper Matsubara self-energy at the Fermi surface.
First, we extrapolate the real part of $\Sigma({\bf k},m)$ to the $m=-1/2$ limit using parabolic
fits with respect to $m$ to obtain $\Sigma'({\bf k})$. We then solve numerically
the equation $\epsilon ({\bf k})-\mu + \Sigma'({\bf k})=0$,
where $\epsilon ({\bf k})=-2[\cos (k_x) + \cos (k_y)] $ is the bare tight-binding dispersion relation,
to determine the shape of the Fermi surface (FS) in the interacting system.
Similarly, for any point on the FS, we obtain the quasiparticle residue by extrapolating data
for $b({\bf k},m)=-\Sigma''({\bf k},m)/\omega_m$ to the $m=-1/2$ limit; according to the FL theory,
$Z({\bf k}) =[1+b({\bf k})]^{-1}$ where $b = - \lim_{\omega \to 0} \partial \Sigma({\bf k},\omega)/\partial \omega$.
Next, the FS velocity is obtained by taking the gradient along the normal direction to FS:
$v_F ({\bf k}_F) = Z({\bf k}_F) \nabla_{\perp}[ \epsilon ({\bf k})-\mu +\Sigma'({\bf k})]_{k\in FS}$.
Finally, the effective mass renormalization, $m_0/m^*$, is deduced by dividing $v_F ({\bf k}_F)$ by the
corresponding FS velocity of the non-interacting gas at the same density.  Except for the largest
density, we find that the anisotropy in $Z$ and $m^*/m_0$ is very small.

\begin{figure}[tb]
\includegraphics[scale=0.4,width=0.9\columnwidth]{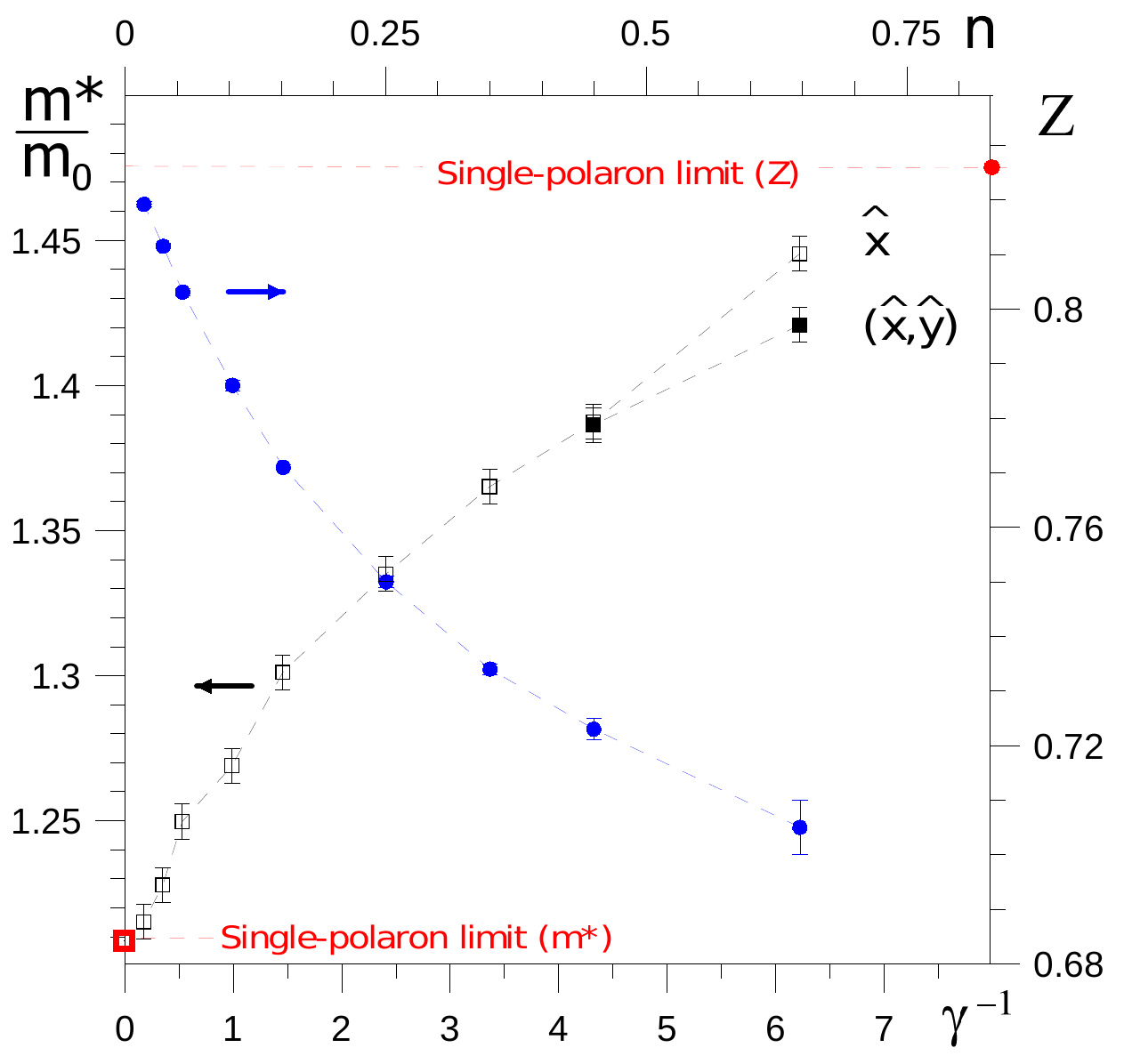}
\caption{\label{fig:4} (color online) Effective mass along
$\hat{x}$ (black open squares) and $(\hat{x},\hat{y})$ (black filled squares) directions
and the quasiparticle residue (blue circles) as functions of the adiabatic parameter
at $\lambda=0.45$ deduced from the same set of simulations as in Fig.~\ref{fig:3}.
The upper horizontal axis provides an approximate density scale.
The error bars for $Z$ at large $\gamma^{-1}$
are not statistical; they indicate the anisotropic spread of $Z$ values on the FS.
The anisotropy of the effective mass is significant only for the largest density, while for all
other points it is unmeasurably small.
}
\end{figure}

Our results for $Z$ and $m^*/m_0$ are shown in Fig.~\ref{fig:4}. Contrary to expectations that
for a single polaron vertex corrections are the strongest and lead to increased renormalization of
quasiparticle properties the data unambiguously indicates that particles are more heavily
``dressed" in the FL regime.
[Further proof that our FL type analysis is correct and simulation temperatures are low enough
for this analysis to be valid is provided by excellent agreement with the single-polaron ($T=0$) results
in the low-density limit.] We interpret these results as follows: scattering restrictions imposed
by the Pauli principle do not overcome the increased low-energy phase space
available for scattering at finite, as opposed to zero, particle momenta on the FS
(the density of states alone can only partially account for this effect at large densities),
not to mention that higher-order terms also admit dressing by particle-hole pairs.

The effective mass renormalization mostly follows $Z$ because
$Z m^*/m_0 $ is approximately constant and close to unity over the entire density range,
reflecting weak momentum dependence of the self-energy in Holstein model, as discussed above
in relation to the insets in Fig.~\ref{fig:1} and Fig.~\ref{fig:2}.

To conclude, we established that the BDMC technique provides
an effective method for solving the many-polaron (or generic interacting electron-phonon) problem
with high and controlled accuracy. The skeleton series converge for moderate values of the dimensionless
coupling $\lambda$ even in the non-adiabatic parameter regime,
and final results for FL parameters can be obtained with sub percent accuracy
after vertex corrections are accounted for.
We find that the quasiparticle ``dressing" is enhanced in the FL regime
relative to the single-polaron case. We verified that the local self-energy assumption is an accurate
(at 2\% level for Holstein model) approximation used in the DMFT and momentum average methods.
Future work should aim at adding electron-electron interactions into the picture,
studies of the phonon spectrum renormalization, superconducting instability, etc.

We thank B. Svistunov for discussions. This work was supported by
the Simons Collaboration on the Many Electron Problem,
the National Science Foundation under the grant PHY-1314735, and
the MURI Program ``New Quantum Phases of Matter" from AFOSR.
N.N. is supported from  Grant-in-Aids for Scientific Research (Kiban S, No. 24224009) from the
Ministry of Education, Culture, Sports, Science and Technology (MEXT).


\end{document}